# Generation of 1.5-octave intense infrared pulses by nonlinear interactions in DAST crystal

C. Vicario[1]*, B. Monoszlai[2], G. Arisholm[3], and C.P. Hauri[1,4]

[1]Paul Scherrer Institut, SwissFEL, 5232 Villigen PSI, Switzerland

[2]ELI-ALPS, ELI-Hu Nkft., Szeged, Hungary and Institute of Physics and University of Pécs Hungary 7624 Pécs Ifjúság út 6

[3]FFI (Norwegian Defence Research Establishment), P.O. Box 25, NO-2027 Kjeller, Norway

[4]Ecole Polytechnique Federale de Lausanne, 1015 Lausanne, Switzerland

*Corresponding author: carlo.vicario@psi.ch

**ABSTRACT**

Infrared pulses with large spectral width extending from 1.2 to 3.4 µm are generated in the organic crystal DAST (4-N, N-dimethylamino-4'-N'-methylstilbazolium tosylate). The input pulse has a central wavelength of 1.5 µm and 65 fs duration. With 2.8 mJ input energy we obtained up to 700 µJ in the broadened spectrum. The output can be easily scaled up in energy by increasing the crystal size together with the energy and the beam size of the pump. The ultra-broad spectrum is ascribed to cascaded second order processes mediated by the exceptionally large effective $\chi^2$ nonlinearity of DAST, but the shape of the spectrum indicates that a delayed $\chi^3$ process may also be involved. Numerical simulations reproduce the experimental results qualitatively and provide an insight in the mechanisms underlying the asymmetric spectral broadening.

**Introduction**

Energetic broadband laser pulses in the short wavelength infrared region (SWIR, 1-3 µm) have attracted interest in the recent years due to their numerous applications ranging from high harmonic generation (HHG) [1], multidimensional vibrational spectroscopy [2], and hyperspectral imaging in medicine [3] to solar cell inspection and imaging applications. Over the last years various generation techniques have been explored, and up to a few hundred µJ few-cycle pulses were generated by self-phase modulation or filamentation in gas or by cross-polarized wave frequency mixing XPW [4-7]. However, the generation of high power pulses and ultra-broad spectra is still challenging [8, 9].

Large spectral broadening has been demonstrated in phase-mismatched second harmonic generation (SHG) in materials with large second order susceptibility [10-13]. Among them the organic crystal DAST (4-N, N-dimethylamino-4'-N'-methylstilbazolium tosylate) [14] was recently investigated for its exceptionally high nonlinear coefficient. Two-cycle pulses with spectra ranging from 1.2 to 2.4 µm were recorded from DAST driven by a short SWIR pump [15]. Organic crystals have also attracted large attention for their potential to generate several MV/cm single-cycle electric fields at THz frequencies by optical rectification [16-18]. In this context, moderate red-shift on the optical pump was observed in DSTMS (4-N,N-dimethylamino-4'-N'-methyl-stilbazolium 2,4,6-trimethylbenzenesulfonate) due to cascaded optical rectification associated with efficient THz emission [19].

The experiment reported here builds on reference [15]. We demonstrate that pumping the DAST crystal at 1.5 µm with the optimal pump fluence allows extending the spectrum towards infrared wavelengths up to 3.4 μm, limited by strong absorption in the organic material. In order to have more pump energy available we have used the 65 fs pump pulse from the optical parametric amplifier (OPA) directly instead of first broadening it in a filament as in ref [15]. We obtain much larger broadening and SWIR energy up to 0.7 mJ per pulse, corresponding to energy conversion efficiency of 25%. The process reported here is intrinsically scalable up to several mJ SWIR energy by using large aperture crystals and higher pump energy.

**Methods**

The compact experimental setup is shown in Fig. 1. A DAST crystal with 200 μm thickness and an aperture of 5 mm is illuminated by a collimated laser beam at 1.5 μm. The pulses originate from a high-energy OPA directly pumped by a 20 mJ, 50 fs Ti:Sa laser at a repetition rate of 100 Hz [21]. The OPA consists of white light generation stage followed by three parametric amplifiers using BBO as nonlinear medium. The system delivers pulses with duration of 65 fs FWHM, maximum energy of 2.8 mJ and intensity stability of 1.6% rms. The pump beam was approximatively Gaussian with diameter of 5 mm at $e^{-2}$ of the maximum. For realizing the maximum spectral broadening the DAST is orientated with the a-axis parallel to the pump polarization to take advantage of the largest nonlinear coefficient: $d_{11}$ = 290 pm/V. This is the typical crystal orientation chosen for efficient THz generation by optical rectification. Second harmonic generation with the orthogonal polarization would be less efficient because of smaller $d_{eff}$ and greater phase mismatch.

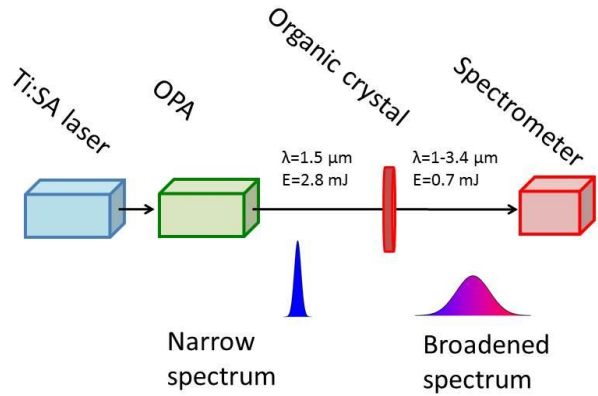

*Fig. 1. Overview of the experimental setup. An optical parametric amplifier (OPA) pumped by a multi-mJ femtosecond Ti:Sa initiates the spectral broadening in DAST. The multi-octave SWIR spectrum is measured with a scanning near infrared spectrometer.*

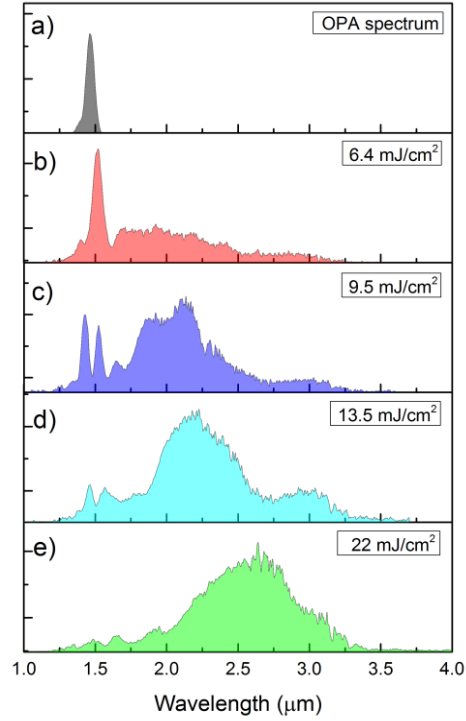

*Fig. 2. Experimental results on spectral broadening in 200 μm DAST for different values of the peak fluence. As illustrated the original spectrum of the OPA (a) is progressively broadened and red-shifted towards SWIR (b-e). The spectral intensity is shown on a linear scale.*

The spectrum generated in DAST is measured using a near-infrared spectrum analyzer based on a tunable acousto-optic bandpass filter [22]. The acousto-optic interaction takes place in a 25 mm long $TeO_2$ crystal. Scanning the frequency of a monochromatic acoustic wave allows selection of a narrow optical spectral line. A Peltier-cooled HgCdTe detector records the intensity of the diffracted optical beam at maximum repetition rate of 100 Hz. The overall spectral intensity is then reconstructed by post-processing based on integration and normalization. The spectrometer has sensitivity over the wavelength range between 1 and 5 μm with resolution better than 5 $cm^{-1}$ and a dynamic range larger than 40 dB. In the experiment, the beam was focused at the spectrometer entrance slit with typical fluence of 60-130 μJ/$cm^2$. The extremely broad SWIR spectrum generated in the experiment makes direct temporal characterization unfeasible [23].

**Experimental results**

Shown in Fig. 2 are the spectra generated in the DAST crystal pumped with different peak fluences. The original OPA spectrum shown in grey (Fig.2a) is 80 nm wide at FWHM and centered at 1500 nm. As the pump intensity is increased the spectrum is progressively broadened and shifted toward SWIR.

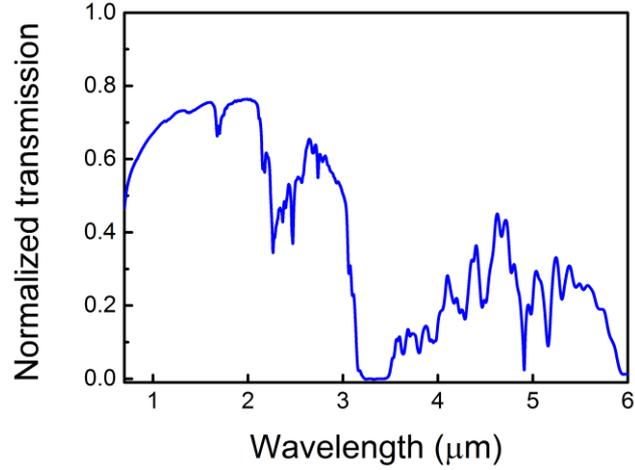

*Fig. 3. Linear unpolarized spectral transmission of a 200 μm thick DAST sample measured by Fourier transform spectroscopy. The curve has not been corrected for the Fresnel losses at the input and output surfaces, which are approximately 13% per surface for the DAST refractive index of 2.1.*

For 6.4 mJ/cm² pump a long tail towards the long wavelengths is visible beside the original spectrum (Fig. 2b). At higher pump fluence, more energy is converted from the pump to near-infrared frequencies (Fig. 2c-2d). Finally at the highest pump fluence of 22 mJ/cm² (green curve), close to the optical damage threshold of DAST, the pump is completely converted to a supercontinuum with central wavelength at 2.6 μm (Fig. 2e). At this fluence the output spectrum covers a continuous range from 1.2 to 3.4 μm, which corresponds to about 1.5 octaves. The low frequency cut-off can be ascribed to the strong absorption in DAST. Fig. 3 shows the linear transmission spectrum of the 200 μm DAST sample.

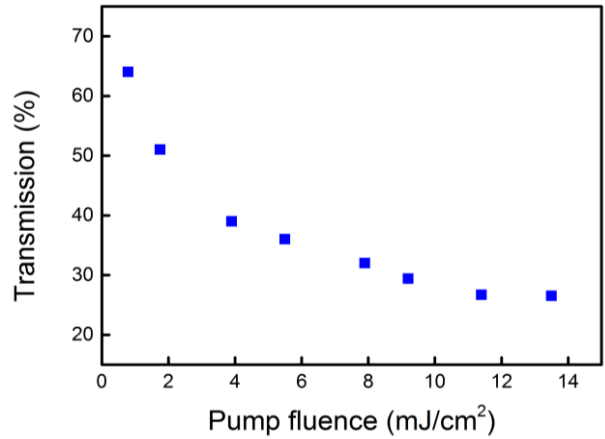

*Fig. 4 Total transmission in DAST pumped at 1500 nm, including the Fresnel losses, as function of the peak input fluence.*

Another interesting feature is the nonlinear transmission as function of the pump fluence shown in Fig. 4. At low input fluences the transmission (not corrected for Fresnel losses), is as high as 73%, while as the pump fluence increases it drops rapidly to less than 40% and saturates at just below 30 % for pump fluence larger than 10 mJ/cm².

**Discussion and simulations**

There are two distinct cascaded χ²-processes which can contribute to the spectral broadening: phase mismatched second harmonic generation (SHG) [12] and electro-optical modulation driven by optical rectification [24]. Phase mismatched SHG can gives rise to strong broadening in a process similar to a Kerr-like nonlinearity, with negative sign (self-defocussing) when Δk > 0. Such a fast nonlinearity typically generates a symmetrically broadened spectrum. On the other hand, the optical rectification in DAST produces an intense THz pulse, which co-propagates with the incoming optical beam and modulates the index of refraction by the electro-optical effect [20]. In the experiment for largest spectral broadening the beam is polarized along the crystal axis which gives rise to the highest THz emission and therefore to the maximum electro-optical coupling. The nature of the electro-optical interaction depends on the group indices of the THz pulse and the optical pulse, which are about 2.4 and 2.2, respectively, for our parameters. This gives an interaction length comparable to the length of the DAST crystal. Because the optical pulse propagates slightly faster than the THz pulse, the THz field seen by any point on the optical pulse depends on the preceding part of the optical pulse. This gives rise to an effective delayed nonlinear response, which, as shown in the following, is consistent with the observed asymmetric spectral broadening.

We have performed numerical simulations in order to evaluate how well these two mechanisms explain the observed results and their relative importance. However, because of incomplete knowledge of the optical properties of DAST, our aim is only to simulate the qualitative features of the spectral broadening, not to achieve accurate agreement. The simulations were performed using the nonlinear optical numerical code Sisyfos [25], which was recently generalized to handle mixing processes within a single, wide-band beam, similar to the method reported in [26]. Because the simulation includes the optical beams and the THz beam in the same field, all the second order processes are characterized by a single χ²-element, and there is no separate parameter for the electro-optic coefficient. We neglected the orthogonally polarized THz wave because it is driven by a much smaller χ²-element and has much greater group velocity mismatch. The program can also include linear absorption, nonlinear refraction, two-photon absorption, and diffraction, but the latter feature is not important in the present case, where the beam is wide and the nonlinear crystal is short. The real field is written $E(z,t) = e(z,t) \exp(-i(\omega_0 t - k_0 z)) + c.c.$, where z is the coordinate along the propagation direction, the transverse position arguments have been suppressed, and the central frequency $\omega_0$ is chosen so that the frequency range of the complex amplitude $e$ runs from 0 to $2\omega_0$. The propagation equations for e are solved in (spatial and temporal) frequency space and have the form:

$$\frac{de_j}{dz} = -\frac{\alpha_j}{2} e_j + iK \frac{\omega_j}{n_j} P_j^{NL} \qquad (1)$$

where the index j runs over spatial and temporal frequency modes, α is the absorption coefficient, $\omega_j$ is the angular frequency, $n_j$ is the refractive index, K is a constant, and $P_j^{NL}$ is the nonlinear driving term for mode j. $P^{NL}$ is computed in real space and then transformed to the frequency domain. In order to avoid aliasing, the upper half of the simulation frequency range is removed from $P^{NL}$ before applying the driving term. The frequency range of the simulation must therefore be wide enough to include the 2nd harmonics of all physically significant frequency components.

The refractive index and the absorption coefficient for the DAST are not known over this wide spectral range. We assumed the refractive index from [27] in the range 0 to 10 THz and the Sellmeier equation from [14] above 50 THz. Due to the lack of other information, we simply interpolated the refractive index in the intermediate range. For absorption we used the data shown in Fig. 3, extended it in the THz range with data from [27] and assumed a high absorption in the intermediate range and at higher frequencies. We used a constant $d_{111}$ of 290 pm/V, which has been measured by SHG at 1542 nm [28], but in reality it should vary over the frequency range because of several absorption features [28]. However, it is worth noting that the value calculated from the electro-optic coefficient is similar [29]. We do not have information on higher-order nonlinear effects (two-photon absorption, nonlinear refraction and Raman effect) at the relevant wavelength.

In order to understand the predominant mechanism, the nonlinear effects are considered separately and in combined together. Figure 5 shows the spectra obtained from simulations where the 1500 nm pump beam was a plane wave with peak fluence 13.5 mJ/cm² and a Gaussian pulse shape with 65 fs (FWHM) duration, as in the experiment. The temporal resolution was 0.5 fs, but we also tried different resolutions and checked the results for consistency. The step length in the propagation direction was adjusted adaptively by the solver for the differential equations. Mismatched SHG alone (a) gives an almost symmetric broadening, similar to self-phase modulation. On the other hand, the optical rectification and electro-optic modulation (OR-EO) mechanism (b) produces a spectrum which is mostly broadened to the red side because of the effective delayed response caused by the slower THz wave. The strong modulations, which are not seen in the experimental results, can be ascribed to the plane-wave simulation. The positions of the maxima and minima depend on pump intensity, so we expect them to be smoothed out for a pump beam with realistic transverse intensity variations.

The spectrum from mismatched SHG and OR-EO together (g) is also broadened mostly towards the red side, but compared to the experimental data it has too much energy in wavelengths shorter than 1.5 μm and too little around 3 μm. A nonlinear refraction process with delayed response, such as Raman or thermal effects, can contribute to asymmetric spectral broadening and red-shift. In order to test this hypothesis we added a delayed nonlinear index term proportional to the integrated intensity:

$$\Delta n(t) = a \cdot \int_0^t I(\tau) d\tau \qquad (2)$$

with a = 8 ×10⁻⁴ m²/J. This is intended as an approximate representation of some nonlinear index mechanism with a long time constant compared to the pulse length. The resulting spectrum, show in Fig. 5 (d), is in better qualitative agreement with the experiments. For completeness we also show the spectra resulting from the delayed nonlinear index alone (c) and in combination with OR-EO (e) or mismatched SHG (f). Mismatched SHG appears to be of relatively small importance, but since this is the mechanism with the least uncertain parameters there is no reason to omit it.

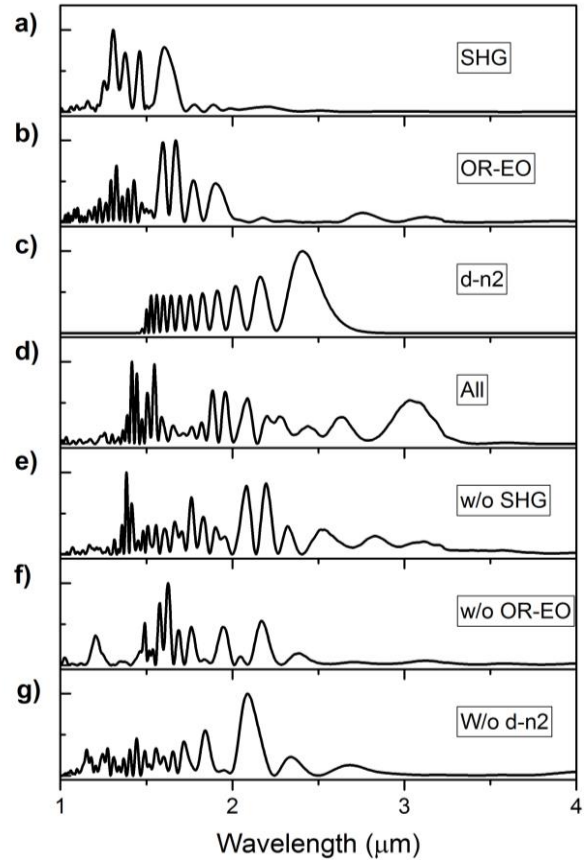

*Fig. 5 Simulations indicate that mismatched SHG, electro-optical modulation due to optical rectification (OR-EO) and delayed nonlinear refraction (d-n2) contribute to the spectral broadening. a)-c) single process do not mimic the experimental spectrum. d) Large broadening and redshift similar to the experiment are observed when all the three processes are taken into account. e)-g) If one of the nonlinear mechanisms is omitted the agreement with the experimental data becomes worse. The simulations are carried out for a plane wave pump of fluence of 13. 5 mJ/cm$^2$.*

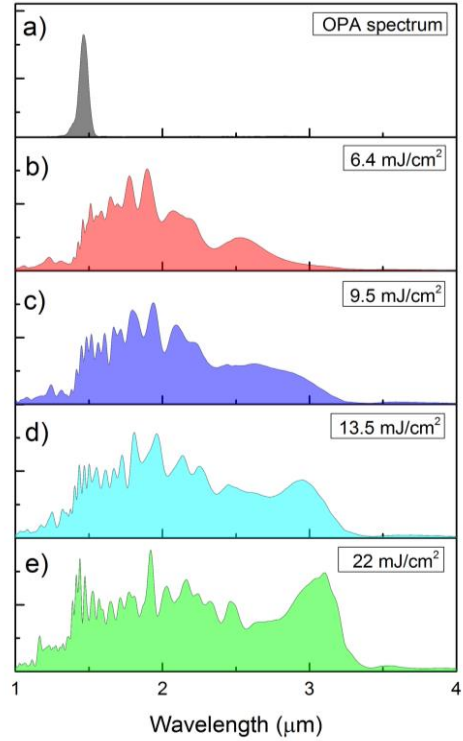

*Fig. 6. Simulated spectral broadening in DAST for the same pump fluence as reported in Fig. 2.*

By shifting photons to lower frequencies, the delayed index of refraction mechanism also increases the effective absorption. Thus, even without explicitly including nonlinear absorption, the transmission is reduced to about 30% at the highest fluence, in fair agreement with the nonlinear absorption data shown in Fig 4.

Figure 6 shows spectra from simulations with all three mechanisms included and Gaussian pump beams of different values of peak fluence, corresponding to those in the experiments. As expected the oscillations in the spectra are smaller than in the plane-wave case. As the pump fluence is increased the output spectrum becomes broader and more shifted toward the long wavelength cutoff of DAST. The qualitative agreement with the measurements of Fig. 2 across the entire spectral range indicates that the observed asymmetric broadening can be explained by cascaded $\chi^2$ and nonlinear refraction processes with delayed response.

The spectral phase and the pulse duration were also extracted from the simulations. Beside the linear and second order terms, the phase-variation in the spectrally broadened pulse is dominated by rapid oscillations, which strongly depend on the pump intensity. For this reason, pulse compression appears to be challenging. For comparison, we have also simulated the pulse duration from a previous experiment [15], which used a shorter pump pulse. In this case the spectral phase oscillations were much reduced, and the numerically compressed pulses were consistent with the two-cycle pulse obtained in the experiment. This gives confidence in the simulation model, and it suggests that compression may be facilitated by short pump pulses, for which the fast nonlinearities are relatively more important than delayed effects. Further investigations are needed to fully understand the extreme spectral broadening in the DAST crystal and the limits to the temporal pulse compression.

**Conclusion**

In conclusion we report unprecedented spectral broadening by multiple nonlinear processes in DAST pumped by few mJ, 65 femtosecond pulses at a central wavelength of 1.5 µm. A 1.5-octave spectrum between 1.2 and 3.5 µm is achieved for pump fluence larger than 10 mJ/cm$^2$. The compact experimental setup provides pulse energy up to 700 µJ and conversion efficiency of 25 %. Scaling the pulse energy up to several mJ is feasible with a larger DAST crystal and higher pump energy while keeping its fluence constant. The results are qualitatively reproduced by numerical simulations, which provide an explanation to the spectral broadening process and accounts for the asymmetric spectral broadening towards longer wavelength. Further studies are ongoing to investigate the temporal compression of the multi-octave spectra presented here.

**Acknowledgments**

CPH acknowledges funding from the Swiss National Science Foundation under grant PP00P2_128493 and association to the National Center of Competence in Research (NCCR-MUST).